\documentclass{article}
\usepackage{spconf,amsmath,graphicx,hyperref}
\usepackage{tabularx, multicol, multirow, booktabs}
\usepackage{xcolor}
\usepackage{setspace}
\usepackage{comment}
\title{Partially Edited Audio: Generation, Defense, and Open Problems}
\name{
 \begin{tabular}{@{}c@{}}
 Lin Zhang$^{1}$ \qquad 
You Zhang$^{2}$ \qquad 
Nicholas Evans$^{3}$ \qquad 
Xin Wang$^{4}$ \qquad 
David Harwath$^{5}$ \qquad \\
 Abhinav Dhall$^{6}$ \qquad 
Bowen Shi$^{1}$ \qquad 
Chengzhe Sun$^{7}$ \qquad 
Siwei Lyu$^{7}$ \qquad 
Sanjeev Khudanpur$^{8}$
\end{tabular}
}
\address{
$^{1}$Independent Researcher \qquad
$^{2}$Télécom Paris \qquad
$^{3}$EURECOM \qquad
$^{4}$National Institute of Informatics \\
$^{5}$The University of Texas at Austin \qquad
$^{6}$Monash University  \\
$^{7}$University at Buffalo, State University of New York \qquad
$^{8}$Johns Hopkins University
}

\begin{document}
\ninept
%\doublespacing
%
\maketitle
\begin{abstract}

Recent advances in generative audio models facilitate increasingly flexible and selective editing of linguistic content and paralinguistic, extralinguistic, and other attributes.
This paradigm, named \textit{partially edited audio}, differs from full audio generation as only portions of the real recording are edited. This technique can serve both beneficial and harmful applications. 
We therefore present a concise survey from both generation and defense perspectives, focusing on audio, as a contribution to the ICASSP 2027 special session entitled ``Partially Edited Media: Perspectives from Synthesis and Defense.'' We first define partially edited audio and argue that what counts as ``fake'' or harmful is an application-dependent judgment, not a fixed property of the edit. We then survey audio editing by editing target, approaches, and evaluation, and then defense methods by detection, localization, diarization, and interpretability, summarizing representative benchmarks and toolkits. Finally, we discuss future directions and open problems.
% , including generation-defense co-development and co-evolution, proactive defense, realistic compositional scenarios, and generalization towards other types of audio and other media modalities more broadly. 

\end{abstract}
\begin{keywords}
audio editing, partially edited audio, generative artificial intelligence, audio forensics, generation and defense
\end{keywords}
%

% \section{scope}
% SS: Partially Edited Media: Perspectives from Synthesis and Defense

% neural audio editing, slightly mention general audio editing (DSP, tampering) -> foundation model-based audio editing.
% local/global 
% Global edits operate on the entire audio or its overall characteristics, including background change, foreground change, and global alteration. Depending on the instruction, each sample may involve either a single operation or an arbitrary composition of multiple operations. --- MMAE

\section{Introduction}
\label{sec:intro}

Generative artificial intelligence (GenAI) has made it possible to locally, spatially, or temporally edit different types of media: images can be edited~\cite{ye2026imgedit}, video segments can be regenerated~\cite{VideoPainter}, or text can be polished co-edited by humans and machines~\cite{CoEdIT}, in each case producing a result where unaltered and edited portions coexist. Audio is no exception~\cite{AudioEditingSurvey}.
Capabilities in the latter have been driven by rapid progress in generative audio models. Beyond synthesizing speech from text, recent audio editing systems can be used to substitute, insert, or delete only selected portions of a given recording~\cite{VoiceBox,VoiceCraft}.
%while keeping the rest unchanged, and 

Audio editing introduces a fundamentally different problem to full audio generation. For fully generated audio, the entire signal is model-generated and homogeneous, and it is typically treated as fake from the defense perspective. In contrast, audio editing involves the fine-grained, local modification of specific regions or attributes, while the remaining context is left unchanged~\cite{AudioEditingSurvey}. Audio editing is not limited to altering what is said, but ranges from linguistic content to paralinguistic, extralinguistic, and other attributes. 
For example, the editing of ``\textit{I won the election}'' to ``\textit{I stole the election}'' fundamentally changes the meaning of the utterance even though the majority of the content is unchanged. Similarly, the emotion of an utterance such as ``\textit{I'm fine}'' could be edited to express sadness or threat without changes to the linguistic content.

% These capabilities have lowered the barrier to 
Such audio editing~\cite{AudioEditingSurvey} serves many harmless or beneficial applications such as the correction of mispronounced words and the post-production editing of podcasts. 
Meanwhile, there are also potentially harmful or nefarious applications when misused, such as spreading misinformation, the manipulation of elections, fraud, extortion, reputational damage, or breaching security systems. 
From the defense perspective, such manipulations are less perceptually salient, more difficult to detect, and potentially more harmful in security-sensitive scenarios. For example, partial editing may be used to create deepfakes~\cite{PartialSpoof, PartialEdit}, thereby resulting in more severe and potentially irreversible consequences.

Despite the rapid progress in neural audio editing and the challenges in detection and defense, research efforts have been modest at best and remain somewhat siloed across the audio generation and security communities. Recent works survey speech- and audio-editing tasks and foundation-model-based methods from a generation perspective~\cite{AudioEditingSurvey, SpeechEditingSummary}. One survey has explored fake localization from the perspective of defense against partially edited speech~\cite{LocalizationSurvey}. However, no single framework connects editing utility and forensic accountability across both generation and defense perspectives. 

This paper surveys partially edited audio from two complementary perspectives. We first define partially edited audio and argue that what counts as fake or harmful is an application-dependent judgment, not a fixed property of the edit (\S~\ref{sec:definition}). We then survey it from two perspectives:
From the generation perspective, we reviewed editing targets, approaches, and evaluation. From the defense perspective, we reviewed detection detection, localization, and diarization~\cite{linPhD2024}.
% (\textit{whether} audio content was edited), localization (\textit{where} the editing occurs), and diarization (\textit{what} editing process was used)~\cite{linPhD2024}.
Finally, we discuss future directions and open problems.
% given they are connected by mutual benefit rather than by conflict: the artifacts generation leaves behind are the cues defense relies on, 

% while the verification defense provides can in turn guide higher-quality editing. Based on this connection, we discuss the requirements for trustworthy audio editing.
% Although the broader problem of partially edited media spans text, images, video, and multimodal content as well, we deliberately focus on the audio modality, where synthesis and defense have advanced rapidly and in parallel over the past several years; we revisit how the resulting framework may generalize across modalities in \S~\ref{sec:challenges}.

\section{From Full Generation to Partial Editing}
\label{sec:definition}

\subsection{Definition}
\label{sec:definition_char}

% The terms 
The definition of \textit{partially edited audio} can be based on two dimensions.
The partial/full distinction concerns how much of the signal or how many audio attributes are manipulated. 
The editing/generation distinction concerns whether the process starts from an existing recording and whether selected portions of that recording are preserved.
% Finally, the spoof/deepfake distinction concerns the forensic or deceptive status of the resulting audio. 
% Consequently, a partial spoof does not necessarily involve neural editing, while a neural editing system may generate a new waveform for the edited region and preserve the remaining regions from the original recording. 
For editing, the same operation that serves a legitimate purpose in one setting can, in another, alter a security- or trust-relevant property of the utterance without disclosure. This observation motivates the accountability view, which separates the technical question of whether an edit occurred from the application-dependent question of whether it is harmful.

The two perspectives of this paper cover different but overlapping scopes: the generation perspective (\S~\ref{sec:synthesis}) concerns audio editing in general, whereas the defense perspective (\S~\ref{sec:defense}) concerns defending against partially edited fake speech.

\subsection{Characteristics}
Three characteristics distinguish partially edited audio from related problems. First, \textit{coexistence}: unaltered and edited portions coexist in the same recording, so utterance-level authenticity labels are insufficient. Second, \textit{locality}: an edit can be local along two distinct axes. Temporally, an edited portion can range from a single phoneme to a long conversation; structurally, editing can target a specific linguistic component (e.g., a word) or a global attribute applied over a selected portion (e.g., a prosody, emotion, or acoustic scenes change). Neither the temporal extent nor the structural target is fixed; both are themselves objects of study, and we return to them in \S~\ref{sec:synthesis}. Third, \textit{dual intent}: the same underlying operation may be performed for a legitimate purpose, such as correcting a mispronunciation, or alternatively for malicious purposes, such as the ``I stole the election'' example introduced in \S~\ref{sec:intro}. These characteristics motivate treating partially edited audio as a joint generation-and-defense problem rather than as an incremental extension of either full speech generation or conventional deepfake detection.

\subsection{What counts as ``Fake''? Connection between generation and defense}
\label{sec:fake}

The generation and defense perspectives are not symmetric. Editing is a broadly capable tool (\S~\ref{sec:synthesis}), and nothing about the operation itself determines whether the result is desirable or harmful; only edits that are undisclosed to a party who relies on the recording, and that violate that party's trust assumptions in a way that causes harm, need to be caught. A disclosed prosody correction in a podcast and an undisclosed identity edit in a fraud call are both partially edited audio, but only the latter is something a defense system should flag.  This is also why defense research has so far covered a narrower part of the editing space than generation, concentrating on content edits in speech (Section~\ref{sec:defense}).

Whether an edit should be flagged also depends on which property of the audio an application seeks to verify. Codec-based neural audio editing often reconstructs the entire audio, both content-edited and surrounding content-unchanged segments undergo decoding and re-encoding by the same codec-processing. Whether such content-unchanged but signal-modified segments should be flagged depends on the application: when the application requires locating which segments retain the original recording signal or only concerns the edited content. This distinction should guide training annotations and evaluation protocols so that detectors target the modifications relevant to the intended application~\cite{PartialEdit}.

\textit{What counts as ``harmful'' or ``fake'' is therefore an application-dependent judgment, not a fixed property of the edit}: it depends on disclosure, consent, and the trust assumptions of whoever relies on the recording, none of which a detector can observe from the waveform alone. 
% A potential technical solution would be to detect audio editing as part of the information for decision, but the decision factors also include others such as detecting misinformation~\cite{spmis_misinfo}.

The generation and defense are nevertheless closely related. The artifacts that an editing approach leaves behind, such as boundary discontinuities at concatenation points, can be the cues on which defense relies~\cite{LiuHowDo2024}, and editing approaches supply the data on which detectors are trained and evaluated. In turn, defense provides a safeguard when audio editing was misused. Analyses of detector behavior may also highlight artifacts that warrant attention in editing systems. Whether such insights can be translated into reliable feedback for improving editing quality remains an open problem.

As generation advances, the artifacts it leaves are likely to become fewer and subtler, so reactive detection alone is likely to become less reliable. Hence, a more reliable detector needs to advance. One could argue that this is like an adversarial game, like Generative Adversarial Network (GAN), where the generator is trained to fool a discriminator. However, legitimate editing aims at utility rather than evading detection, and defense aims at verifiability rather than rejecting edits. Verification then increasingly depends on information that the editing system itself embeds at generation time, such as watermarks, which we discuss as future work in \S~\ref{sec:challenges}.

In the following sections, we present a taxonomy of partially edited audio from generation and defense perspectives, separately.
\section{Taxonomy of Partially Edited Audio Generation}
\label{sec:synthesis}

\subsection{Editing Targets -- \textit{What} are Edited?}

The first dimension concerns what is edited. \textit{Content} editing changes the linguistic sequence through word or phrase substitution, insertion, deletion, or mispronunciation correction~\cite{VoiceCraft,EditSpeech}. \textit{Prosody and speaking-style} editing modifies pitch, duration, rhythm, stress, speaking rate, or style~\cite{Morrison2024}. \textit{Paralinguistic} editing, such as emotion editing, changes affective attributes including arousal, valence, or emotional style~\cite{EmoCorrector}. \textit{Acoustic-environment} editing changes music, background noise, or acoustic scenes~\cite{SeamlessEdit}. Beyond speech, editing targets include music, e.g., changing instruments or stems from text prompts or instructions~\cite{MusicMagus,InstructMusicGen}, and general sound, e.g., adding effects or editing individual sound events~\cite{AUDIT,Recomposer}. \textit{Cross-modal editing} further requires synchronization between audio, video, text, or other modalities~\cite{ThinkSound}. Within this cross-modal setting, \textit{object-level editing} narrows the target further, to the sound produced by a specific visual object rather than the full audio-visual scene~\cite{ObjectAVEdit}.

A recent survey of foundation-model-based audio editing categorizes methods along acoustic/semantic/instance axes~\cite{AudioEditingSurvey}, another survey explores text-based speech editing methods~\cite{SpeechEditingSummary}.
% , and a massive multitask benchmark MMAE~\cite{MMAE} evaluates a single instruction-based audio editing across dozens of such editing operations. This benchmark supports treating attribute control as a conditioning mechanism rather than an independent editing type: content, prosody, emotion, music, environment, and multimodal information can all serve as control variables across what is increasingly a shared underlying editing task.

\subsection{Editing Approaches - \textit{How} to edit?}

Approaches differ in how the edited portion is produced and how it is integrated with the unedited context, which in turn determines the artifacts left for defense. We illustrate them with speech, where most approaches were developed; similar patterns hold for music and general audio. The earliest approach is splicing and copy-move: authentic material is cut, inserted, deleted, or copied within a recording by digital signal processing, which gives explicit control but leaves boundary discontinuities and mismatches in background or electrical-network signatures, the very cues that classical tampering forensics exploits~\cite{AudioSplicingEnv}. Generate-then-splice approaches replace the material with speech synthesized or converted by a neural model but still splice it in, so the content is machine-generated while the boundary artifacts remain~\cite{PartialSpoof,LiuHowDo2024}. Context-aware infilling removes this separation by generating the edited portion conditioned on both its left and right context~\cite{EditSpeech}: autoregressive codec language models predict acoustic tokens to fill the masked span~\cite{VoiceCraft}, while non-autoregressive diffusion and flow-matching models infill in spectrogram or latent space~\cite{VoiceBox}. Most recently, a single instruction-conditioned model selects its editing behavior from natural-language prompts across editing targets~\cite{AudioEditingSurvey, AuK}.

% These approaches were largely designed and validated for a single short edits to long-form multi-turn editing, complicated by the crucial need for multiple edits to remain consistent across an utterance or a multi-speaker conversation, is comparatively underexplored~\cite{FluentEditor2}. 
Across these approaches, the major challenge for editing is to jointly optimize edit utility and contextual consistency. 
% Transcription accuracy alone is insufficient as a main solution measure the editing quality

\subsection{Generation-Oriented Evaluation and Resources}
\label{sec:synthesis_eval}

Editing systems should be evaluated from both the edited-portion and unchanged-portion perspectives~\cite{AudioEditingSurvey,MMAE}. Important criteria include content or semantic correctness, speaker similarity, prosodic and emotional consistency, acoustic scenes and background preservation, boundary smoothness, perceptual naturalness, controllability, diversity, and robustness to different editing locations and durations.

Word error rate is insufficient because an edit may achieve the desired transcript while introducing speaker, prosodic, or acoustic inconsistencies. Naturalness is commonly estimated with automatic mean-opinion-score predictors such as UTMOS~\cite{UTMOS}, while speaker consistency between edited and unedited portions is typically measured with embedding-based similarity derived from speaker verification models such as WavLM~\cite{WavLM}. Realistic editing benchmarks such as RealEdit~\cite{VoiceCraft} facilitate evaluation under diverse speakers, recording conditions, background noise, and music, and massive multitask benchmarks such as MMAE~\cite{MMAE} further broaden coverage across editing operations, with an associated ICASSP 2027 challenge\footnote{\url{https://audio-editing-challenge.github.io/}} further encouraging public benchmarking.

\section{Taxonomy of Defense Against Partially Edited Audio}
\label{sec:defense}

% \NZ{Neil: This section seems to contain only speech. I think it's worth mentioning. Lin: good point, thanks! I added an initial long version below}

\S~\ref{sec:synthesis} surveyed audio editing across editing targets and approaches. Defense research covers a narrower part of this space: it has concentrated almost exclusively on speech, mostly on temporally edited utterances whose linguistic content, and hence meaning, has been altered. Its datasets have historically been built by embedding generated portions into bona fide speech rather than by running neural editing models~\cite{PartialSpoof}. This narrower scope follows where trust violations are most consequential (\S~\ref{sec:fake}) and the history of available datasets, and it leaves partially edited deepfake speech, where the two perspectives meet, as a part of the defense problem that has only recently begun to be addressed~\cite{huang2024detecting, PartialEdit}.

Although editing can in principle be applied to any type of audio, including speech, music, and general sound events, the security community currently focuses on speech. This narrow focus reflects the current state of the community rather than an inherent limitation of the underlying problem. We survey defenses against partially edited speech, but we note that the same defense principles, i.e., detection, localization, diarization, and interpretability introduced in this section, could in principle be applied to any type of audio.
% other audio types, such as sound events, music or singing voices.

% Although editing can in principle be applied to any type of audio, including speech, music, and general sound events, existing defense research has concentrated almost exclusively on speech, aiming to detect and localize temporally edits that alter linguistic content and, with it, the meaning of an utterance. This reflects the current focus of the community rather than an inherent limitation of the underlying problem. In the following we survey defenses against partially edited deepfake speech, but note that the same defense principles, i.e., detection, localization, diarization, and interpretability could in principle be applied to other audio types, such as sound events, music or singing voices.

\subsection{Detection and Localization}

Utterance-level deepfake detection answers \textit{whether} an audio recording is authentic or fake. It is used for content screening, access control, and large-scale moderation. However, the authentic portions of a partially edited recording may mislead the decision, especially when the edited portion is small~\cite{PartialSpoof,HAD}. 
Classical splicing detectors exploit low-level physical cues, such as environmental acoustic signatures~\cite{AudioSplicingEnv}. Since utterance-level detection is a nominally similar task to detecting fully deepfake audio, the model architectures are largely shared between the two settings, ranging from conventional machine learning model such as GMM, and deep learning like LCNN-~\cite{LCNN} to self-supervised-learning-based detectors~\cite{SSLCM} and, most recently, LLM-based models~\cite{UnifyingDetectLoc,HoliAntiSpoof}. What matters more for the detection of partially edited deepfake than the choice of architecture is training data: a detector trained only on fully generated utterances degrades substantially on partially edited audio, whereas training directly on partially edited data with short embedded fake portions generalizes reliably to both settings~\cite{PartialSpoof,PartialSpoofTASLP}.

Segment-level deepfake localization answers \textit{when} a deepfake occurs, and can be extended to answer \textit{where} it occurs, whether at the temporal, attribute/component, or object level. Existing methods focus on temporal localization, with most approaches relying on frame-level cues~\cite{PartialSpoofTASLP}, boundary information~\cite{BAM, CFPRF}, or adjacent-frame inconsistency~\cite{PET}.
% frame-level boundary detection~\cite{BDR}, boundary-aware attention~\cite{BAM}, coarse-to-fine proposal refinement~\cite{CFPRF}, boundary-frame cross graph attention~\cite{BFCNet}, high-frequency temporal self-consistency learning~\cite{PET}, and embedding-based temporal localization~\cite{TDL}.
More can be found from a survey focus on localization literature~\cite{LocalizationSurvey}. Most recently, a preliminary study examined whether large language models (LLMs) can be used to localize manipulated words in transcribed partially fake speech~\cite{LLMLocalize}, and another study fine-tuned the large-scale Whisper-based speech recognition model for the same task~\cite{tran2026deepfake}. Other studies have applied LLMs to unify detection and content localization within a single generative framework~\cite{UnifyingDetectLoc,HoliAntiSpoof}.

Besides, a parallel line of work extends detection and localization to the audio-visual and multimodal setting~\cite{AVDeepfake1Mpp,MFMS}.
% , where an edited audio segment is evaluated jointly with corresponding visual or cross-modal cues~\cite{AVDeepfake1Mpp,MFMS}. 

\subsection{Deepfake Diarization}

Detection and localization determine whether and where an edit occurred, but not which generation process produced it, or whether multiple edited regions in the same recording originate from different sources. Deepfake diarization~\cite{SpoofDiarization} extends localization by determining \textit{what} editing process was used when. It can enable fine-grained traceability by linking edited regions to the underlying algorithms and, potentially, to the parties responsible for them.
% It traces the attack source of edited regions by utilizing deepfake localization predictions to enhance the diarization performance. 
% It was further improved by adding token-based attractors and cross-attention mechanisms~\cite{TokenAttractors}. 
This form of fine-grained source tracing remains comparatively underexplored, particularly for longer and more complex recordings, e.g., multi-speaker and long-form audios containing multiple edits from different processes.

\subsection{Interpretability and Analyses}

Beyond raw detection performance, increasing attention has been given to how and why detectors succeed or fail on partially edited audio. Artifacts at edit boundaries can unintentionally serve as detection cues~\cite{Negroni2024}. A systematic study of the cues used by detectors on partially deepfake audio also shows that models rely more on boundary artifacts at concatenation points~\cite{LiuHowDo2024}. These findings are important for understanding failure modes and building trust in detection and localization systems.

\subsection{Defense-Oriented Evaluation and Resources}

Progress in this area has been driven by a growing number of dedicated datasets, challenges, and toolkits. PartialSpoof~\cite{PartialSpoof} and HAD~\cite{HAD} were among the first resources for partially edited deepfake speech detectors for English and Mandarin, respectively, 
%to target needed in partially edited deepfake speech detector, 
followed by the ADD challenge series~\cite{ADD2023}. Subsequent resources have extended coverage to segment-level localization~\cite{Psynd}, audio-visual and multimodal forgery~\cite{AVDeepfake1Mpp, LAVDF}, LLM-driven and instruction-based partial fakes~\cite{LlamaPartialSpoof,PartialEdit}, long-form noisy conditions~\cite{LENSDF}. Complementing these datasets, open-source toolkits such as WeDefense~\cite{WeDefense} and DeepFense~\cite{DeepFense} are emerging, lowering the barrier to reproducible research and benchmarking.

% \input{tables/synthesis_defense}

% \zlin{may consider to move Table~\ref{tab:unified-view} to here if more space avalible.}

% Throughout this paper we accordingly organize both perspectives around three informal notions: whether an edit achieves its intended \textit{utility}, whether it stays \textit{consistent} with its unedited context, and whether it can be verified for \textit{accountability} when its legitimacy needs to be judged. Accountability, in particular, is a \textit{verifiability} requirement, i.e., the ability to answer ``was this region edited, where, and by what process,'' rather than a requirement to classify every edit as fake; the legitimacy judgment itself is left to the receiving application. This is why defense research has concentrated overwhelmingly on content and speaker-identity edits (\S~\ref{sec:defense}): these edit types are most likely to violate a high-stakes trust assumption, not because other edit types are inherently safe to leave unverified. We formalize these three notions as an evaluation framework in \S~\ref{sec:challenges}.

% Table~\ref{tab:representative_work} in Appendix~\ref{app:table} summarizes representative studies organized along these lines.

\section{Future Directions and Open Problems}
\label{sec:challenges}

\begin{comment}
    
\textbf{Co-Evaluation from Generation and Defense Perspectives:} 
With the development of neural editing methods, it is likely that newer generation methods reduce the artifacts that past detection methods rely on, as we discussed in \S~\ref{sec:fake}.
We call for generation research to report not only editing performance improvement, but also how much it fails the defense systems as a side effect. Rather than as separate research threads, we believe this will benefit both communities. 
\end{comment}

\textbf{Co-Development of Generation and Defense:}
As discussed in \S~\ref{sec:fake}, generation and defense are closely related, yet the two communities progress separately. Each advance in editing defines a new defense problem: the shift from splicing to context-aware infilling degrades the performance of models that have not encountered such edits before~\cite{PartialEdit}. Moreover, matched editing systems can supply the data needed to train and evaluate corresponding defenses. Conversely, defense makes editing more deployable. Reliable detection is part of what enables powerful editing systems to be released responsibly; Voicebox~\cite{VoiceBox}, for instance, was withheld from public release because of concerns about misuse. The two perspectives can therefore progress together: for example, by releasing samples edited by new systems for defense evaluation, as ASVspoof has been incorporating newly developed systems into successive evaluations.

\textbf{Proactive Protection:}
We also claim that trustworthy editing should consider utility-consistency-accountability: whether an edit achieves its intended \textit{utility}, whether it stays \textit{consistent} with its unedited context, and whether it can be verified for \textit{accountability} when authenticity or manipulation must be transparent. Accountability can be achieved proactively rather than only reactively: watermarks can be embedded at generation time so that subsequent edits can be identified or flagged~\cite{AudioSeal}. Alternatively, watermarking authentic recordings at the point of capture establishes a provenance record before any editing occurs.
Since foundation models increasingly underlie both editing and detection systems, such mechanisms need to remain robust against further edits, re-encoding, and adversarial removal.

% \textbf{Human-Centered and Trustworthy Audio Editing:} Studies should connect model decisions with human perception and downstream risks. Evaluation should also address the fairness, privacy, and consent implications of voice biometric systems affected by partial editing~\cite{FairnessVoiceBiometrics,BenchmarkBiasPrivacy}, as well as the prevention of misuse and the responsible release of powerful editing systems.

\textbf{Toward Realistic Scenarios:} Most existing generation and defense approaches are evaluated under simplified conditions that might not be representative of audio captured in more realistic, practical scenarios.
%real-world audio rarely satisfies. 
Current benchmarks contain mostly only clean, single-speaker, studio-quality recordings with single, short edits; in practice, however, audio recordings contain  background noise, reverberation, compression, and potentially multiple edits. 
% for which realistic benchmarks such as 
Benchmarks are also overwhelmingly monolingual, and multilingual editing remains unexplored. Finally, realistic conversational audio may contain overlapping speakers, multiple edited portions attributable to different sources, and edits at different semantic levels simultaneously (e.g., word-level content changes combined with utterance-level style transfer). How to extend both generation and defense methods to fully multi-speaker, multi-object settings is still an open problem.

\textbf{Generalization and Domain Adaptation:}
Generalization is a common challenge in deepfake defense.
Detectors are usually trained and evaluated in-domain, and their performance can degrade sharply under domain shift. Existing datasets already differ in attack types, languages, and recording conditions, which can further degrade performance. 
Partially edited audio is especially fragile in the face of domain variability because the shifts are set by the generation side. For example, detectors trained on spliced partial spoofs tend to exploit boundary artifacts~\cite{LiuHowDo2024}, which context-aware neural editing largely mitigated. 
In addition, partially edited audio introduces another dimension: editing style. For example, in most existing benchmarks, fake audio is created through agent-driven editing. Studies have shown that models trained on one editing style may perform poorly when applied to another~\cite{LLMLocalize}.

Domain adaptation has been developed primarily to defend against fully generated deepfake audio, but it has received little attention in the context of partially edited audio, for both detection and localization. How to build a robust detector for partially edited audio therefore remains an open problem.

% Promising directions include adaptation from weak utterance-level labels, few-shot adaptation to newly released editing systems, and training with diverse partially edited samples, which improved out-of-domain performance whereas additional fully synthetic or genuine data degraded it.

\textbf{Defense Against Editing on Other Audio Types, Attributes, and Modalities:} 
Current defense research has focused on almost exclusively upon linguistic content editing. Advances in GenAI increasingly facilitate the selective manipulation of paralinguistic, extralinguistic, and other attributes: changes to intonation or emotional expression~\cite{EmoCorrector} can alter the perceived intent of an utterance, while replacing background sounds~\cite{ESDD2} can change its apparent context, even when the spoken words remain unchanged. For example, editing the intonation of a voicemail to sound more urgent or distressed, or replacing a call's background noise to imply a different location, can mislead a listener's judgment.
Such perceptual manipulations have been explored in other modalities~\cite{gupta2025multiverse}, but remain underexplored in audio. 
Moreover, as noted at the beginning of \S~\ref{sec:defense}, existing defenses primarily address the editing of speech. Extending detection, localization, diarization, and interpretability to music, singing voice~\cite{zhang2024svdd}, and general sound editing requires further attention. 

% The problem framing developed in this paper is not unique to audio, either: P
% Partially edited content is emerging as a broader challenge across text, images, video, and multimodal media, in which only selected portions are altered. Early audio-visual work already extends localization and detection beyond the audio-only setting~\cite{LAVDF,AVDeepfake1M}.
% suggesting that core defense principles developed for partially edited deepfake audio are conceptually transferable across modalities. 
% An open direction is whether the utility--consistency--accountability view introduced can be generalized into a modality-agnostic accountability framework.

\S~\ref{sec:intro} opened by noting that images, video, and text, not only audio, can be locally edited so that unaltered and edited portions coexist. Generation for these other modalities is advancing just as quickly as audio editing. Early audio-visual work extends localization and detection beyond the audio-only setting~\cite{LAVDF,AVDeepfake1M}, but the problems specific to the multimodal setting itself, such as maintaining cross-modal consistency and synchronization between audio, video, and text (\S~\ref{sec:synthesis}), and defense against them from a defense perspective, remain open on both the generation and defense sides.

\section{Conclusions}
\label{sec:conclusion}

Partially edited audio contains both unaltered and edited portions in one recording. We defined it, discussed its characteristics, and examined the connection between the generation and defense perspectives, noting that \textit{editing has dual intent and that what counts as ``fake'' is an application-dependent judgment rather than a fixed property of the edit}. 

Reviewing both perspectives, our survey shows that generation increasingly covers speech, music, and general sound, and can edit linguistic, paralinguistic, extralinguistic, and other attributes, whereas defense remains concentrated on temporally edited linguistic content in speech. Yet the two are closely related: the artifacts that editing leaves behind are the cues on which defense relies, and legitimate editing seeks utility, not evasion. We therefore argue that trustworthy editing requires utility, contextual consistency, and accountability, i.e., the ability to detect, localize, and trace an edit when trust is at stake. As edits grow more seamless, accountability should be designed into editing systems through co-development and proactive provenance rather than left to reactive detection alone. How to build trustworthy audio editing, and how to defend against increasingly flexible editing, remain open problems that neither community can solve alone.

% We presented a survey of partially edited audio according to editing targets, mechanisms, and evaluation protocols. We further proposed a unified utility-consistency-accountability view, emphasizing that practical audio editing should be useful, contextually consistent, and appropriately verifiable. Future progress will require closer cooperation between generation and defense communities, particularly for neural editing, long-form conversations, proactive protection, and trustworthy audio editing systems. It will likely need to engage with the broader multimodal landscape of partially edited media as that field continues to develop.


\bibliographystyle{IEEEbib}
\let\oldthebibliography\thebibliography
\let\endoldthebibliography\endthebibliography
\renewenvironment{thebibliography}[1]{%
  \oldthebibliography{#1}%
  \setlength{\itemsep}{1.5pt}%
  \setlength{\parsep}{0pt}%
  \setlength{\parskip}{0pt}%
}{\endoldthebibliography}
{\fontsize{8}{8.6}\selectfont
\bibliography{main}}

\end{document}